\documentclass[journal]{IEEEtran}

\pdfoutput=1
\usepackage{enumitem}
\usepackage{amsmath, amssymb}
\usepackage[pdftex]{graphicx}
\graphicspath{{./Figures/}{../Figures/}}
\usepackage{epstopdf}
\usepackage{float}
\usepackage{subcaption}
\usepackage{booktabs}
\usepackage{threeparttable}
\usepackage{tablefootnote}
\usepackage{amsthm}
    
    \newtheorem{assumption}{Assumption}
    \newtheorem{definition}{Definition}
    
    \newtheorem{remark}{Remark}
    \newtheorem{theorem}{Theorem}

    \newtheorem{proposition}{Proposition}

\usepackage[ruled, lined, ,longend, linesnumbered]{algorithm2e}

\usepackage[hang,flushmargin]{footmisc}
\usepackage{algpseudocode}
\usepackage{enumitem}
\SetKwInOut{Input}{Input}
\SetKwInOut{Declares}{Declares}
\SetKwInOut{Require}{Require}
\SetKw{KwBy}{by}
\SetKw{KwAnd}{and}
\SetKw{KwEndFor}{end for}

\usepackage{fancyhdr}

\DeclareFontFamily{OT1}{pzc}{} 
\DeclareFontShape{OT1}{pzc}{m}{it}{<-> s * [1.10] pzcmi7t}{}
\DeclareMathAlphabet{\mathpzc}{OT1}{pzc}{m}{it}

\newcommand{\cc}[1]{{\mathcal{#1}}} 
\def\R{ {\rm \,I\!R} } 
\def\Z{{\mathbb{Z}}} 
\def\inR#1{\in\R^{#1}} 
\def\vv#1{{ \rm \bf{#1}}} 
\newcommand{\T}{^\top} 
\def\fracg#1#2{{\displaystyle{\frac{#1}{#2}}}} 
\def\sp#1#2{\langle #1,#2\rangle } 
\def\Sum#1#2{\sum\limits_{#1}^{#2}} 
\DeclareMathSymbol{\shortminus}{\mathbin}{AMSa}{"39} 

\def\bmat#1{\left[\begin{array}{#1}} 
\def\emat{\end{array}\right]} 
\newcommand {\bsis} {\left\{ \begin{array} }
\newcommand {\esis} {\end{array}\right.}
\newcommand{\bRlist}{\renewcommand{\labelenumi}{(\roman{enumi})} \begin{enumerate}} 
\newcommand{\eRlist}{\end{enumerate} \renewcommand{\labelenumi}{\arabic{enumi}}} 

\def\tx{\tilde{x}}
\def\tu{\tilde{u}}

\begin{document}
\pagestyle{fancy}

\title{Implementation of model predictive control for tracking in embedded systems using a sparse extended ADMM algorithm%
}

\author{Pablo~Krupa,~Ignacio~Alvarado,~Daniel~Limon,~Teodoro~Alamo%
    \thanks{Department of Systems Engineering and Automation, Universidad de Sevilla, 41092, Seville, Spain (e-mails: \texttt{pkrupa@us.es}, \texttt{ialvarado@us.es} \texttt{dml@us.es}, \texttt{talamo@us.es}). Corresponding author: Pablo Krupa.}%
    \thanks{This work was supported in part by the Agencia Estatal de Investigación (AEI) under Grant PID2019-106212RB-C41/AEI/10.13039/501100011033, in part by the MCIU-Spain and FSE under Grant FPI-2017, and in part by the Junta de Andalucía and the Fondos Europeos para el Desarrollo Regional (FEDER) under Grant P20\_00546.}
\thanks{Accepted version of the article published in IEEE Transactions on Control Systems Technology. DOI:~10.1109/TCST.2021.3128824~\hfill}
\thanks{\copyright{}2021 IEEE. Personal use of this material is permitted.  Permission from IEEE must be obtained for all other uses, in any current or future media, including reprinting/republishing this material for advertising or promotional purposes, creating new collective works, for resale or redistribution to servers or lists, or reuse of any copyrighted component of this work in other works.~\hfill}%
}

\maketitle
\thispagestyle{fancy}

\begin{abstract}
    This article presents a sparse, low-memory footprint optimization algorithm for the implementation of the model predictive control (MPC) for tracking formulation in embedded systems.
    This MPC formulation has several advantages over standard MPC formulations, such as an increased domain of attraction and guaranteed recursive feasibility even in the event of a sudden reference change. However, this comes at the expense of the addition of a small amount of decision variables to the MPC's optimization problem that complicates the structure of its matrices.
    We propose a sparse optimization algorithm, based on an extension of the alternating direction method of multipliers, that exploits the structure of this particular MPC formulation.
    We describe the controller formulation and detail how its structure is exploited by means of the aforementioned optimization algorithm.
    We show closed-loop simulations comparing the proposed solver against other solvers and approaches from the literature.
\end{abstract}


\section{Introduction}

Model Predictive Control (MPC) is an advanced control strategy in which the control action is obtained, at each sample time, from the solution of an optimization problem where a prediction model is used to forecast the evolution of the system over a finite prediction horizon. One of the main advantages of MPC over other control strategies is that it inherently considers and satisfies system constraints \cite{Camacho_S_2013}.

There are many different MPC formulations in the literature, each of which is defined by an optimization problem with different objective function and/or set of constraints \cite{Rawlings_MPC_2017}. In general, the optimization problem is posed as a minimization problem in which the objective function penalizes the distance between the reference and the predicted system evolution over the prediction horizon. In this paper we focus on linear MPC controllers for tracking whose optimization problem can be posed as a quadratic programming (QP) problem.

Since an MPC controller requires solving an optimization problem at each sample time, its use has historically been confined to computationally powerful devices, such as PCs. However, as evidenced by the references provided below, there is a growing interest in the literature in the implementation of these controllers in devices with very limited computational and memory resources, known as \textit{embedded systems}.

One possible approach for implementing MPC in embedded systems is to use \textit{explicit} MPC \cite{Tondel_A_2003}, in which the solution of the MPC optimization problem is stored as a lookup table that is computed offline. However, this lookup table can become prohibitively large for medium to large-sized systems and/or for MPC problems with many constraints. Some examples of it being implemented in embedded systems are \cite{Valencia_IETCTA_2012, Velagic_2014_ExplicitMPC}.

Another approach comes from the recent development of optimization solvers for QP problems that are tailored to embedded systems. A few of the most widespread ones include OSQP \cite{Stellato_OSQP}, qpOASES \cite{Ferreau_2014_qpOASES}, CVXGEN \cite{CVXGEN}, FiOrdOs \cite{FiOrdOs} and FORCES \cite{FORCES}.
These solvers, although used to successfully implement MPC controllers in embedded systems (see \cite{Huyck_MED_2012, Binder_CCA_2015, Hartley_IETCST_2014, Kufoalor_MED_2014} for a few examples), are for generic QP problems.
Therefore, the development of optimization algorithms tailored to the specific MPC optimization problem can potentially provide better results.
Some examples of solvers that fall into this category include PRESAS \cite{Quirynen_OCAM_2020}, FalcOpt \cite{FalcOpt}, HPMPC \cite{Frison_ECC_2014} or $\mu$AO-MPC \cite{Zometa_2013_muAOMPC}.
Other noteworthy examples of MPC-tailored solvers implemented in embedded systems include \cite{Jerez_IETAC_2014, Sabo_CDC_2018, Shukla_SD_2017} for implementations in FPGAs, \cite{Currie_M2VIP_2012} for microcontrollers and \cite{Krupa_TCST_20, Krupa_ECC_18, Lucia_IETII_2018, Pereira_MED_2015, Necoara_ECC_2013} for programmable logic controllers.
Specifically, the authors proposed in \cite{Krupa_TCST_20} and \cite{Krupa_ECC_18} a method for the embedded implementation of MPC controllers which takes advantage of the structure of the matrices of the QP problem.
This led to the development of a sparse optimization algorithm, based either on the FISTA \cite{Beck_SJIS_2009} or on the alternating direction method of multipliers (ADMM) \cite{Boyd_FTML_2011} algorithms.

The references on embedded MPC provided above, as well as most publications in this field, deal with standard MPC formulations, presenting efficient ways to solve them.
In this paper, however, we are concerned with providing an efficient solver for a more intricate linear MPC formulation known as \textit{MPC for tracking} (MPCT) \cite{Ferramosca_A_2009}, which has several advantages over standard MPC formulations that makes it especially attractive for its implementation in embedded systems.
For instance, it offers a significantly larger domain of attraction than standard MPC formulations with the same prediction horizon, which is useful in embedded systems since the use of smaller prediction horizons reduces the computational and memory footprint of the controller.
Additionally, it provides recursive feasibility even in the event of a sudden reference change \cite{Limon_A_2008}.

However, these advantages come at the cost of a slightly more complicated QP problem due to the inclusion of additional decision variables.
In this paper we propose a sparse solver for MPCT based on the \textit{extended} ADMM (EADMM) algorithm \cite{Cai_EADMM_2017}.
The use of this algorithm leads to sparse matrices whose simple structure can be efficiently exploited to attain a computationally and memory efficient implementation following a similar approach to the one used in \cite{Krupa_TCST_20}, but for a more intricate MPC formulation.
In particular, we avoid representing the sparse matrices using the typical sparse-matrix representations (e.g., \textit{compressed sparse column}).

We showcase the benefits of the proposed solver by comparing it to other ADMM-based approaches to control a simulated ball and plate system.
Additionally, the results of its implementation in a Rapsberry Pi to control a two-wheeled inverted pendulum robot have been reported in \cite{Krupa_ECC_21_arXiv}, providing further evidence of its good performance.

The proposed solver is part of the Matlab toolbox ``SPCIES: Suite of Predictive Controllers for Industrial Embedded Systems'' \cite{Spcies}, available at \url{https://github.com/GepocUS/Spcies}.

The remainder of this paper is structured as follows.
Section \ref{sec:problem:formulation} describes the problem formulation and control objective.
The MPC for tracking formulation is described in Section \ref{sec:MPCT}.
The extended ADMM algorithm is detailed in Section \ref{sec:EADMM}.
Section \ref{sec:solving:MPCT} describes how the MPCT optimization problem is recast into a problem solvable by the EADMM algorithm. 
Section \ref{sec:numerical:results} shows the numerical results.
Finally, conclusions are drawn in Section \ref{sec:conclusions}.

\subsubsection*{Notation}

Given two vectors $x, y \inR{n}$, $x \leq (\geq) \; y$ denotes componentwise inequalities and $\sp{x}{y}$ is their standard inner product.
For a vector $x\inR{n}$ and a positive definite matrix $A\inR{n \times n}$, $\|x\| \doteq \sqrt{\sp{x}{x}}$, $\|x\|_A \doteq \sqrt{\sp{x}{A x}}$ is its weighted Euclidean norm, and $\| x \|_\infty \doteq \max_{i = 1 \dots n}{| x_{(i)} |}$, where $x_{(i)}$ is the $i$-th element of $x$, is its $\ell_\infty$-norm.
For a symmetric matrix $A$, $\| A \|$ denotes its spectral norm.
Given scalars and/or matrices $M_1, M_2, \dots, M_N$ (not necessarily of the same dimensions), we denote by $\texttt{diag}(M_1, M_2, \dots, M_N)$ the block diagonal matrix formed by the diagonal concatenation of $M_1$ to $M_N$.
Given a matrix $A \inR{n \times m}$, $A_{i,j}$ denotes its $(i,  j)$-th element, $A\T$ its transposed and $A^{-1}$ its inverse (if it is non-singular).
$(x_{1}, x_{2}, \dots, x_{N})$ is a column vector formed by the concatenation of column vectors $x_{1}$ to $x_{N}$.
Given two integers $i$ and $j$ with ${j \geq i}$, $\Z_i^j$ denotes the set of integer numbers from $i$ to $j$, i.e. ${\Z_i^j \doteq \{i, i+1, \dots, j-1, j\}}$. 

\begin{definition}
A function $f: \R^n \rightarrow \R$ is said to be $\mu$-strongly convex, for some scalar $\mu > 0$, if $f(x) - \frac{\mu}{2} \| x \|^2$ is convex.
\end{definition}

\section{Problem formulation} \label{sec:problem:formulation}

We consider a controllable system described by a discrete linear time-invariant state-space model%
\begin{equation} \label{eq:model}
    x_{k+1} = A x_k + B u_k,
\end{equation}
where $x_k \inR{n}$ and $u_k \inR{m}$ are the state and input of the system at sample time $k$, respectively. Additionally, we consider that the system is subject to the box constraints
\begin{subequations} \label{eq:constraints}
\begin{align}
    &\underline{x} \leq x_k \leq \overline{x}, \\
    &\underline{u} \leq u_k \leq \overline{u}.
\end{align}
\end{subequations}

The control objective is to steer the system to the given reference $(x_r, u_r)$ while satisfying the system constraints \eqref{eq:constraints}. This will only be possible if the reference is an admissible steady state of the system, which we formally define as follows.

\begin{definition} \label{def:Admissible}
    An ordered pair $(x_a, u_a) \in \R^n \times \R^m$ is said to be an admissible steady state of system (\ref{eq:model}) subject to (\ref{eq:constraints}) if
    \bRlist
        \item $x_a = A x_a + B u_a$, i.e., it is a steady state of system (\ref{eq:model}),
        \item $\underline{x} \leq x_a \leq \overline{x}$,
        \item $\underline{u} \leq u_a \leq \overline{u}$.
    \eRlist
\end{definition}

If the given reference is not an admissible steady state of system \eqref{eq:model} subject to \eqref{eq:constraints}, then we wish to steer the system to the closest admissible steady state, for some given criterion of closeness.

\section{Model Predictive Control for Tracking} \label{sec:MPCT}

The MPCT formulation \cite{Ferramosca_A_2009} differs from other standard MPC formulations in the inclusion of a pair of decision variables $(x_s, u_s)$ known as the \textit{artificial reference}. The cost function penalizes, on one hand, the difference between the predicted states and control actions with this artificial reference, and on the other, the discrepancy between the artificial reference and the reference $(x_r, u_r)$ given by the user. In particular, this paper focuses on the MPCT formulation shown below, which uses a terminal equality constraint.

For a given prediction horizon $N$, the MPCT control law for a given state $x$ and reference $(x_r, u_r)$ is derived from the solution of the following convex optimization problem%
\begin{subequations} \label{eq:MPCT} 
\begin{align}  
    \min\limits_{\substack{\vv{x}, \vv{u},\\ x_s, u_s}} \;& \Sum{i = 0}{N-1} \left( \| x_i {-} x_s \|^2_Q {+} \| u_i {-} u_s \|^2_R \right) {+} \| x_s {-} x_r \|^2_T {+} \| u_s {-} u_r \|^2_S \\
    s.t.& \; x_0 = x \label{eq:MPCT:initial} \\
        & \; x_{i+1} = A x_i + B u_i, \; i\in\Z_0^{N-1} \label{eq:MPCT:prediction} \\
        & \; \underline{x} \leq x_i \leq \overline{x}, \; i\in\Z_1^{N-1} \\
        & \; \underline{u} \leq u_i \leq \overline{u}, \; i\in\Z_0^{N-1} \\
        & \; x_s = A x_s + B u_s \label{eq:MPCT:steady:state}\\
        & \; \underline{x} + \varepsilon_x \leq x_s \leq \overline{x} - \varepsilon_x \label{eq:MPCT:ineq:x_s} \\
        & \; \underline{u} + \varepsilon_u \leq u_s \leq \overline{u} - \varepsilon_u \label{eq:MPCT:ineq:u_s} \\
        & \; x_N = x_s, \label{eq:MPCT:terminal}
\end{align}
\end{subequations}
where the decision variables are the predicted states and inputs $\vv{x} = ( x_0, \dots, x_{N} )$, $\vv{u} = ( u_0, \dots, u_{N-1} )$ and the artificial reference $(x_s, u_s)$; the positive definite matrices $Q \in \R^{n \times n}$ $R \in \R^{m \times m}$, $T \in \R^{n \times n}$ and $S \in \R^{m \times m}$ are the cost function matrices; and $\varepsilon_x \inR{n}$, $\varepsilon_u \inR{m}$ are arbitrarily small positive vectors which are added to avoid a possible loss of controllability when the constraints are active at the equilibrium point \cite{Limon_A_2008}.

One of the properties of the MPCT formulation \eqref{eq:MPCT} is that it will steer the closed-loop system to the admissible steady state $(x_a, u_a)$ that minimizes the cost ${\| x_a - x_r \|^2_T + \| u_a - u_r \|^2_S}$ \cite{Ferramosca_A_2009, Limon_A_2008}. Additionally, as previously mentioned in the introduction, this formulation provides significantly larger domains of attraction than other standard MPC formulations, and guarantees recursive feasibility even in the event of a sudden reference change.

\section{Extended ADMM} \label{sec:EADMM}

This section introduces the \textit{extended} ADMM algorithm \cite{Cai_EADMM_2017}, which, as its name suggests, is an extension of the ADMM algorithm \cite{Boyd_FTML_2011} to optimization problems with more than two separable functions in the objective function. In particular, we focus on the following class of optimization problem.

Let $\theta_i : \R^{n_i} \rightarrow \R$ for $i \in\Z_1^3$ be convex functions, ${\cc{Z}_i \subseteq \R^{n_i}}$ for $i \in\Z_1^3$ be closed convex sets, $C_i \inR{m_z \times n_i}$ for $i \in\Z_1^3$ and $b \inR{m_z}$.
Consider the optimization problem
\begin{subequations} \label{eq:EADMM:optimization:problem}
\begin{align}
    \min\limits_{z_1, z_2, z_3} &\Sum{i = 1}{3} \theta_i(z_i) \\
    s.t.& \; \Sum{i = 1}{3} C_i z_i = b \\
        & \; z_i \in \cc{Z}_i, \; i\in\Z_1^3.
\end{align}
\end{subequations}
and let its augmented Lagrangian $\cc{L}_\rho(z_1, z_2, z_3, \lambda)$ be given by
\begin{equation} \label{eq:lagrangian}
    \cc{L}_\rho(\cdot) = \Sum{i = 1}{3} \theta_i(z_i) + \langle \lambda, \Sum{i = 1}{3} C_i z_i {-} b \rangle + \frac{\rho}{2} \left\| \, \Sum{i = 1}{3} C_i z_i {-} b \, \right\|^2,
\end{equation}
where $\lambda \inR{m_z}$ are the dual variables and $\rho > 0$ is the penalty parameter. We denote a solution point of \eqref{eq:EADMM:optimization:problem} by $(z_1^*, z_2^*, z_3^*, \lambda^*)$, assuming that one exists.

Algorithm \ref{alg:EADMM} shows the implementation of the extended ADMM algorithm for a given exit tolerance $\epsilon > 0$ and initial points $(z_2^0, z_3^0, \lambda^0)$. The superscript $k$ indicates the value of the variable at iteration $k$.
We note that step \ref{alg:EADMM:step:exit} uses the $\ell_\infty$-norm, although any other norm can be used.
Algorithm \ref{alg:EADMM} returns an $\epsilon$-suboptimal solution $(\tilde z_1^*, \tilde z_2^*, \tilde z_3^*, \tilde \lambda^*)$ of problem \eqref{eq:EADMM:optimization:problem}. As shown in \cite{Chen_EADMM_convergence_2016}, the EADMM algorithm is not necessarily convergent under the typical assumptions of the classical ADMM algorithm. However, multiple results have shown its convergence under additional assumptions \cite{Cai_EADMM_2017, Chen_EADMM_convergence_2013, Lin_EADMM_convergence_2015} or by adding additional steps \cite{He_EADMM_2012, Li_EADMM_2015}. In particular, \cite{Cai_EADMM_2017} proved its convergence under the following assumption, as stated in the following theorem.

\begin{assumption}[\cite{Cai_EADMM_2017}, Assumption 3.1] \label{ass:EADMM}
    Functions $\theta_1$ and $\theta_2$ are convex; function $\theta_3$ is $\mu_3$-strongly convex for some $\mu_3 > 0$; and $C_1$ and $C_2$ are full column rank.
\end{assumption}

\begin{theorem}[Convergence of EADMM;~\cite{Cai_EADMM_2017},~Theorem~3.1] \label{theo:convergence:EADMM}
    Suppose that Assumption \ref{ass:EADMM} holds and that the penalty parameter ${\rho \in \left( 0, \, \frac{6 \mu_3}{17 \| C_3\T C_3 \|} \right)}$. Then, the sequence of points $(z_1^k, z_2^k, z_3^k)$ generated by Algorithm \ref{alg:EADMM} converges to a point in the optimal set of problem \eqref{eq:EADMM:optimization:problem} as $k \rightarrow \infty$.
\end{theorem}

\begin{remark} \label{rem:EADMM:exit:cond}
    The exit condition given in step \ref{alg:EADMM:step:exit} of Algorithm~\ref{alg:EADMM} serves as an indicator of the (sub-)optimality of the current iterate \cite[\S 5]{Cai_EADMM_2017}.
\end{remark}

\begin{algorithm}[t]
    \DontPrintSemicolon
    \caption{Extended ADMM} \label{alg:EADMM}
    \Require{$z_2^0$, $z_3^0$, $\lambda^0$, $\rho > 0$, $\epsilon > 0$ \label{alg:EADMM:exit:cond}}
    $k \gets 0$\;
    \Repeat{$\| \Gamma \|_\infty {\leq} \epsilon$, $\| z_2^{k} {-} z_2^{k-1}\|_\infty {\leq} \epsilon$, $\| z_3^{k} {-} z_3^{k-1}\|_\infty {\leq} \epsilon$ \label{alg:EADMM:step:exit}}{
        $z_1^{k+1} {\gets} \arg\min\limits_{z_1} \left\{ \cc{L}_\rho(z_1, z_2^k, z_3^k, \lambda^k) \; | \; z_1 {\in} \cc{Z}_1 \right\}$\; \label{alg:EADMM:step:z_1}
    $z_2^{k+1} {\gets} \arg\min\limits_{z_2} \left\{ \cc{L}_\rho(z_1^{k+1}, z_2, z_3^k, \lambda^k) \; | \; z_2 {\in} \cc{Z}_2 \right\}$\; \label{alg:EADMM:step:z_2}
        $z_3^{k+1} {\gets} \arg\min\limits_{z_3} \left\{ \cc{L}_\rho(z_1^{k+1}, z_2^{k+1}, z_3, \lambda^k) \; | \; z_3 {\in} \cc{Z}_3 \right\}$\; \label{alg:EADMM:step:z_3}
        $\Gamma \gets \Sum{i = 1}{3} C_i z_i^{k+1} - b$\; \label{alg:EADMM:step:residual}
        $\lambda^{k+1} \gets \lambda^k + \rho \Gamma $\; \label{alg:EADMM:step:lambda}
        $k \gets k + 1$\;
    }
    \KwOut{$\tilde z_1^* \gets z_1^{k}$, $\tilde z_2^* \gets z_2^{k}$, $\tilde z_3^* \gets z_3^{k}$, $\tilde \lambda^* \gets \lambda^{k}$}
\end{algorithm}

\section{Solving MPCT using extended ADMM} \label{sec:solving:MPCT}

This section describes how problem \eqref{eq:MPCT} is solved using Algorithm \ref{alg:EADMM}. The objective is to develop a memory and computationally efficient algorithm so that it can be implemented in an embedded system. To this end, we recast problem \eqref{eq:MPCT} so that steps \ref{alg:EADMM:step:z_1}, \ref{alg:EADMM:step:z_2} and \ref{alg:EADMM:step:z_3} are easy to solve following a similar approach to the one taken in \cite{Krupa_TCST_20}. Algorithm \ref{alg:EADMM:MPCT} shows the particularization of Algorithm \ref{alg:EADMM} that results from this effort.

\subsection{Recasting the MPCT problem} \label{sec:recast:MPCT}

Let us define $\tx_i \doteq x_i - x_s$ and $\tu_i \doteq u_i - u_s$. Then, we can rewrite \eqref{eq:MPCT} as
\begin{subequations} \label{eq:MPCT:transformed} 
\begin{align}  
    \min\limits_{\substack{ \tilde{\vv{x}}, \tilde{\vv{u}}, \vv{x},\\ \vv{u}, x_s, u_s} } \; &\Sum{i = 0}{N} \left( \| \tx_i \|^2_Q + \| \tu_i \|^2_R \right) + \| x_s - x_r \|^2_T + \| u_s - u_r \|^2_S \label{eq:MPCT:transformed:cost:function} \\
    s.t.& \; x_0 = x \label{eq:MPCT:transformed:initial} \\
        & \; \tx_{i+1} = A \tx_i + B \tu_i, \; i\in\Z_0^{N-1} \label{eq:MPCT:transformed:model} \\
        & \; \underline{x} \leq x_i \leq \overline{x}, \; i\in\Z_1^{N-1} \label{eq:MPCT:transformed:ineq:x} \\
        & \; \underline{u} \leq u_i \leq \overline{u}, \; i\in\Z_0^{N-1} \label{eq:MPCT:transformed:ineq:u}\\
        & \; \underline{x} + \varepsilon_x \leq x_N \leq \overline{x} - \varepsilon_x \label{eq:MPCT:transformed:ineq:x_N} \\
        & \; \underline{u} + \varepsilon_u \leq u_N \leq \overline{u} - \varepsilon_u \label{eq:MPCT:transformed:ineq:u_N}\\
        & \; x_s = A x_s + B u_s \label{eq:MPCT:transformed:xsus}\\
        & \; \tx_i + x_s - x_i = 0, \; i\in\Z_0^{N} \label{eq:MPCT:transformed:x}\\
        & \; \tu_i + u_s - u_i = 0, \; i\in\Z_0^{N} \label{eq:MPCT:transformed:u}\\
        & \; x_N = x_s \label{eq:MPCT:transformed:terminal:x_s} \\
        & \; u_N = u_s, \label{eq:MPCT:transformed:terminal:u_s}
\end{align}
\end{subequations}
where the decision variables are ${\tilde{\vv{x}} = (\tx_0, \dots, \tx_N)}$, ${\tilde{\vv{u}} = (\tu_0, \dots, \tu_N)}$, ${\vv{x} = (x_0, \dots, x_N)}$, ${\vv{u} = (u_0, \dots, u_N)}$, $x_s$ and $u_s$. Equality constraints \eqref{eq:MPCT:transformed:x} and \eqref{eq:MPCT:transformed:u} impose the congruence of the decision variables with the original problem. We note that inequalities \eqref{eq:MPCT:ineq:x_s} and \eqref{eq:MPCT:ineq:u_s} are omitted because they are already imposed by \eqref{eq:MPCT:transformed:ineq:x_N} and \eqref{eq:MPCT:transformed:ineq:u_N} alongside the inclusion of \eqref{eq:MPCT:transformed:terminal:x_s} and \eqref{eq:MPCT:transformed:terminal:u_s}.

Note that the summations in the cost function \eqref{eq:MPCT:transformed:cost:function} now include $i = N$. However, this does not change the solution of the optimization problem due to the inclusion of \eqref{eq:MPCT:transformed:terminal:x_s} and~\eqref{eq:MPCT:transformed:terminal:u_s}.

We can now obtain a problem of form \eqref{eq:EADMM:optimization:problem} by taking
\begin{subequations} \label{eq:MPCT:z:selection}
\begin{align}
    z_1 &= (x_0, u_0, x_1, u_1, \dots, x_{N-1}, u_{N-1}, x_N, u_N), \label{eq:MPCT:z:selection:z1}\\
    z_2 &= (x_s, u_s), \label{eq:MPCT:z:selection:z2} \\
    z_3 &= (\tx_0, \tu_0, \tx_1, \tu_1, \dots, \tx_{N-1}, \tu_{N-1}, \tx_N, \tu_N), \label{eq:MPCT:z:selection:z3}
\end{align}
\end{subequations}
which leads to $\theta_1 (z_1) = 0$,
\begin{align*}
    &\theta_2 (z_2) = \frac{1}{2} z_2\T \texttt{diag}(T, S) z_2 - (T x_r, S u_r)\T z_2, \\
    &\theta_3(z_3) = \frac{1}{2} z_3\T \texttt{diag}(Q, R, Q, R, \dots, Q, R) z_3,
\end{align*}
\begin{align*} 
    &C_1 = \bmat{ccc} [I_n \, 0_{n, m}] & 0 & 0 \\\hline -I_{n+m} & 0  & 0 \\ 0 & \ddots & 0 \\ 0 & 0 & -I_{n+m} \\\hline 0 & 0 & -I_{n+m} \emat,
    && C_2 = \bmat{c} 0 \\\hline I_{n+m} \\ \vdots \\ I_{n+m} \\\hline I_{n+m} \emat, \nonumber \\
    &C_3 = \bmat{ccc} 0 & \dots & 0 \\\hline I_{n+m} & 0  & 0 \\ 0 & \ddots & 0 \\ 0 & 0 & I_{n+m} \\\hline 0 & \dots & 0 \emat,
    && b = \bmat{c} x \\\hline 0 \\ \vdots \\ 0 \\\hline 0 \emat.
\end{align*}

Matrices $C_1$, $C_2$ and $C_3$ contain the equality constraints \eqref{eq:MPCT:transformed:initial}, \eqref{eq:MPCT:transformed:x}, \eqref{eq:MPCT:transformed:u}, \eqref{eq:MPCT:transformed:terminal:x_s} and \eqref{eq:MPCT:transformed:terminal:u_s}. Specifically, the first $n$ rows impose constraint \eqref{eq:MPCT:transformed:initial}, the last $n+m$ rows impose the constraints \eqref{eq:MPCT:transformed:terminal:x_s} and \eqref{eq:MPCT:transformed:terminal:u_s}, and the rest of the rows impose the constraints \eqref{eq:MPCT:transformed:x} and \eqref{eq:MPCT:transformed:u}. Set $\mathcal{Z}_1$ is the set of vectors $z_1$ \eqref{eq:MPCT:z:selection:z1} for which the box constraints \eqref{eq:MPCT:transformed:ineq:x}-\eqref{eq:MPCT:transformed:ineq:u_N} are satisfied; set $\mathcal{Z}_2$ is the set of vectors $z_2$ \eqref{eq:MPCT:z:selection:z2} that satisfy the equality constraint \eqref{eq:MPCT:transformed:xsus}; and set $\mathcal{Z}_3$ is the set of vectors $z_3$ \eqref{eq:MPCT:z:selection:z3} that satisfy the equality constraints \eqref{eq:MPCT:transformed:model}.

We note that our selection of $z_i$ and $C_i$ for $i\in\Z_1^3$ results in an optimization problem that satisfies Assumption~\ref{ass:EADMM}. Therefore, under a proper selection of $\rho$, the iterates of the EADMM algorithm will converge to the optimal solution of the MPCT controller. In practice, the parameter $\rho$ may be selected outside the range shown in Theorem \ref{theo:convergence:EADMM} in order to improve the convergence rate of the algorithm \cite{Cai_EADMM_2017}. In this case, the convergence will not be guaranteed and will have to be extensively checked with simulations.

\subsection{Particularizing EADMM to the MPCT problem} \label{sec:solving:MPCT:EADMM}

By taking $z_i$ and $C_i$ for $i\in\Z_1^3$ as shown in Section \ref{sec:recast:MPCT}, we can particularize Algorithm \ref{alg:EADMM} to the MPCT problem, resulting in Algorithm \ref{alg:EADMM:MPCT}. This algorithm requires solving three QP problems (which we label \ref{eq:QP1}, \ref{eq:QP2} and \ref{eq:QP3} in the following) with explicit solutions at each iteration. The control action to be applied to the system are the elements $u_0$ of the variable $\tilde z_1^*$ \eqref{eq:MPCT:z:selection:z1} returned by Algorithm \ref{alg:EADMM:MPCT}.

Step \ref{alg:EADMM:MPCT:step:z_1} minimizes the Lagrangian \eqref{eq:lagrangian} over $z_1$, resulting in the following box-constrained QP problem,
\begin{align} \label{eq:QP1}
    \cc{P}_1(z_2^k, z_3^k, \lambda^k): \; \min\limits_{z_1} \; &\frac{1}{2}\, z_1\T H_1 z_1 + q_1\T z_1 \tag{$\cc{P}_1$} \\
    s.t. \; & \underline{z}_1 \leq z_1 \leq \overline{z}_1, \nonumber
\end{align}
where $H_1 = \rho C_1\T C_1$,
\begin{align*}
    q_1 &= \rho C_1\T C_2 z_2^k + \rho C_1\T C_3 z_3^k + C_1\T \lambda^k - \rho C_1\T b, \\
    \underline{z}_1 &= (-M_n, \underline{u}, \underline{x}, \dots, \underline{u}, \underline{x} + \varepsilon_x, \underline{u} + \varepsilon_u), \\
    \overline{z}_1 &= (M_n, \overline{u}, \overline{x}, \dots, \overline{u}, \overline{x} - \varepsilon_x, \overline{u} - \varepsilon_u),
\end{align*}
and $M_n \inR{n} > 0$ has arbitrarily large components.

Due to the structure of $C_1$, matrix $H_1$ is a positive definite diagonal matrix. As such, each element $j\in\Z_1^{(N+1) (n+m)}$ of the optimal solution of \ref{eq:QP1}, denoted by $(z_1^*)_j$, can be explicitly computed as,
\begin{equation} \label{eq:solve:QP1}
    (z_1^*)_j = \max \left\{ \min \left\{ \fracg{- (q_1)_j}{(H_1)_{j,j}}, \; (\overline{z}_1)_j \right\}, \; (\underline{z}_1)_j \right\}.
\end{equation}

\begin{remark} \label{rem:constraints}
    In this article we consider box constraints \eqref{eq:constraints} so that $z_1^{k+1}$ is very simple to update, as shown in \eqref{eq:solve:QP1}.
    However, the use of more generic constraints $\underline{y} \leq C x + D u \leq \overline{y}$ would also be possible. 
    In this case, \eqref{eq:solve:QP1} would result in $N+1$ decoupled inequality-constrained QPs, which would have to be solved at each iteration of the algorithm, but that, given their small scale, may be solved rather cheaply using interior point or active set methods.
\end{remark}

Step \ref{alg:EADMM:MPCT:step:z_2} minimizes the Lagrangian \eqref{eq:lagrangian} over $z_2 = (x_s, u_s)$, resulting in the following equality-constrained QP problem,
\begin{align} \label{eq:QP2}
    \cc{P}_2(z_1^{k+1}, z_3^k, \lambda^k): \; \min\limits_{z_2} \; &\frac{1}{2}\, z_2\T H_2 z_2 + q_2\T z_2 \tag{$\cc{P}_2$} \\
    s.t. \; & G_2 z_2 = b_2, \nonumber
\end{align}
where $H_2 = \texttt{diag}(T, S) + \rho C_2\T C_2$, $G_2 = [(A - I_n) \;\; B]$, $b_2 = 0$ and
\begin{equation*}
    q_2 = -(T x_r, S u_r) + \rho C_2\T C_1 z_1^{k+1} + \rho C_2\T C_3 z_3^k + C_2\T \lambda^k - \rho C_2\T b.
\end{equation*}
This problem has an explicit solution derived from the following proposition \cite[\S 10.1.1]{Boyd_ConvexOptimization}.

\begin{proposition} \label{prop:optimality:QP}
    Consider an optimization problem $\min\limits_z \, (1/2) z\T H z + q\T z,\; s.t.\; G z = b$, where $H$ is positive definite. A vector $z^*$ is an optimal solution of this problem if and only if there exists a vector $\mu$ such that,
    \begin{align*}
        &G z^* = b \\
        &H z^* + q + G\T \mu = 0,
    \end{align*}
    which using simple algebra and defining $W_H \doteq G H^{-1} G\T$, leads to
    \begin{subequations} \label{eq:solve:QP}
    \begin{align}
        &W_{H} \mu = -(G H^{-1} q + b) \label{eq:solve:QP:mu} \\
        &z^* = - H^{-1} (G\T \mu + q). \label{eq:solve:QP:z}
    \end{align}
    \end{subequations}
\end{proposition}

The optimal solution $z_2^*$ of problem \eqref{eq:QP2} can be obtained by substituting \eqref{eq:solve:QP:mu} into \eqref{eq:solve:QP:z}, which leads to the expression
\begin{equation} \label{eq:solve:QP2}
    z_2^* = M_2 q_2, 
\end{equation}
where $M_2 = H_2^{-1} G_2\T (G_2 H_2^{-1} G_2\T)^{-1} G_2 H_2^{-1} - H_2^{-1} \inR{(n+m)\times(n+m)}$. This matrix, which has a relatively small dimension, is computed offline and stored in the embedded system. Vector $b_2$ does not appear in the above expression because it is equal to zero.

Step \ref{alg:EADMM:MPCT:step:z_3} minimizes the Lagrangian \eqref{eq:lagrangian} over $z_3$, resulting in the following equality-constrained QP problem,
\begin{align} \label{eq:QP3}
    \cc{P}_3(z_1^{k+1}, z_2^{k+1}, \lambda^k): \; \min\limits_{z_3} \; &\frac{1}{2}\, z_3\T H_3 z_3 + q_3\T z_3 \tag{$\cc{P}_3$} \\
    s.t. \; & G_3 z_3 = b_3, \nonumber
\end{align}
where $H_3 = \texttt{diag}(Q, R, Q, R, \dots, Q, R) + \rho C_3\T C_3$, $b_3 = 0$,
\begin{align*}
    q_3 &= \rho C_3\T C_1 z_1^{k+1} + \rho C_3\T C_2 z_2^{k+1} + C_3\T \lambda^k - \rho C_3\T b, \\
    G_3 &= \bmat{cccccccc}
    A & B & -I_n & 0 & \cdots & \cdots & 0 & 0 \\
    0 & 0 & A    & B & -I_n   & \cdots & 0 & 0 \\
    0 & 0 & 0& \ddots & \ddots   & \ddots & 0 & 0 \\
    0 & 0 & 0    & 0 & A      & B      & -I_n & 0
          \emat.
\end{align*}
This problem can be sparsely solved using the following approach.
Let $W_{H_3} \doteq G_3 H_3^{-1} G_3\T$. Due to the sparse structure of $G_3$ and the fact that $H_3$ is a block diagonal matrix, we have that the Cholesky factorization of $W_{H_3}$, that is, the upper-triangular matrix $W_{H_3, c}$ that satisfies $W_{H_3} = W_{H_3, c}\T W_{H_3, c}$, has the following structure,
\begin{equation} \label{eq:W_c} 
    W_{H_3,c} = \left(
                                   \begin{array}{cccccc}
                                     \beta^{1} & \alpha^{1} & .. & .. & 0 & 0 \\
                                     .. & \beta^{2} & \alpha^{2} & .. & .. & 0 \\
                                     .. & .. & .. & .. & .. & .. \\
                                     0 & .. & .. & .. & \beta^{N-1} & \alpha^{N-1} \\
                                     0 & 0 & .. & .. & .. & \beta^{N} \\
                                   \end{array}
                                 \right),
\end{equation}
where we define the sets of matrices $\mathcal{A} = \{ \alpha^1, \dots, \alpha^{N-1} \}$, ${\alpha^{i} \inR{n \times n}}$; and $\mathcal{B} = \{ \beta^{1}, \dots, \beta^{N} \}$, ${\beta^{i} \inR{n \times n}}$.
Note that the amount of memory required to store the sets of matrices $\mathcal{A}$ and $\mathcal{B}$ grows linearly with the prediction horizon $N$.

Then, \eqref{eq:solve:QP:mu} can be solved by consecutively solving the following two systems of equations that use the auxiliary vector $\hat{\mu}$,
\begin{subequations} \label{eq:solve:QP3:mu}
\begin{align}
    W_{H_3, c}\T \hat{\mu} &= -G_3 H_3^{-1} q_3 \label{eq:solve:QP3:mu:step:1} \\
    W_{H_3, c} \mu &= \hat{\mu},
\end{align}
\end{subequations}
which are easy to solve due to $W_{H_3, c}$ being upper-triangular. In fact, these two systems of equations can be computed sparsely by exploiting the structures of $W_{H_3, c}$, $G_3$ and $H_3^{-1}$ \cite[\S 6]{Krupa_arXiv_MPCT_20_v2}. Finally, the optimal solution $z_3^*$ of \ref{eq:QP3} can be obtained as in \eqref{eq:solve:QP:z},
\begin{equation} \label{eq:solve:QP3:z}
    z_3^* = -H_3^{-1} (G_3\T \mu + q_3),
\end{equation}
which, once again, can be computed sparsely by exploiting the structures of $G_3$ and $H_3^{-1}$.

\begin{algorithm}[t] 
    \DontPrintSemicolon
    \caption{Extended ADMM for MPCT} \label{alg:EADMM:MPCT}
    \Require{$z_2^0$, $z_3^0$, $\lambda^0$, $\rho > 0$, $\epsilon > 0$}
    $k \gets 0$\;
    \Repeat{$\| \Gamma \|_\infty \leq \epsilon$, $\| z_2^{k} {-} z_2^{k-1}\|_\infty \leq \epsilon$, $\| z_3^{k} {-} z_3^{k-1}\|_\infty \leq \epsilon$}{
        Obtain $z_1^{k+1}$ by solving \ref{eq:QP1}$(z_2^k, z_3^k, \lambda^k)$ using \eqref{eq:solve:QP1}\; \label{alg:EADMM:MPCT:step:z_1}
        Obtain $z_2^{k+1}$ by solving \ref{eq:QP2}$(z_1^{k+1}, z_3^k, \lambda^k)$ using \hspace{-0.45em} \eqref{eq:solve:QP2}\; \label{alg:EADMM:MPCT:step:z_2}
        Obtain $z_3^{k+1}$ by solving \ref{eq:QP3}$(z_1^{k+1}, z_2^{k+1}, \lambda^k)$ using \eqref{eq:solve:QP3:mu} and \eqref{eq:solve:QP3:z}.\; \label{alg:EADMM:MPCT:step:z_3}
        $\Gamma \gets \Sum{i = 1}{3} C_i z_i^{k+1} - b$\; \label{alg:EADMM:MPCT:step:residual}
        $\lambda^{k+1} \gets \lambda^k + \rho \Gamma $\; \label{alg:EADMM:MPCT:step:lambda}
        $k \gets k + 1$\;
    }
    \KwOut{$\tilde z_1^* \gets z_1^{k}$, $\tilde z_2^* \gets z_2^{k}$, $\tilde z_3^* \gets z_3^{k}$, $\tilde \lambda^* \gets \lambda^{k}$}
\end{algorithm}

\begin{remark} \label{rem:nonsingularity} 
Because of the controllability of the system, the fact that $\rho > 0$ and the positive definite nature of matrices $Q$, $R$, $T$ and $S$, it is easy to verify that matrices $G_2 H_2^{-1} G_2\T$ and $G_3 H_3^{-1} G_3\T$ are guaranteed to be nonsingular.
\end{remark}

\begin{remark} \label{rem:rho:matrix} 
It has been shown that the performance of ADMM can be significantly improved by having different values of $\rho$ for different constraints \cite[\S 5.2]{Stellato_OSQP}, i.e., by considering $\rho$ as a diagonal positive definite matrix. In particular, we find that, for our problem, the convergence improves significantly if the equality constraints \eqref{eq:MPCT:transformed:initial}, \eqref{eq:MPCT:transformed:terminal:x_s}, \eqref{eq:MPCT:transformed:terminal:u_s}, \eqref{eq:MPCT:transformed:u} for $i = N$, and \eqref{eq:MPCT:transformed:x} for $i = 0$ and $i = N$, are penalized more than the others.
As in many operator splitting methods, the selection of $\rho$ is a mostly open problem which has a significant impact on the performance of the algorithm \cite[\S 4]{Stathopoulos_FTSC_2016}.
We find, however, that penalizing the above constraints $10$ times more than the others is a good starting point.
\end{remark}

\begin{remark} \label{rem:bound:rho} 
    The theoretical upper bound for $\rho$ provided in Theorem \ref{theo:convergence:EADMM} is easily computable in this case. Indeed, we have that $C_3\T C_3$ is the identity matrix, and therefore its spectral norm $\| C_3\T C_3 \| = 1$. Furthermore, $\mu_3$ is the minimum eigenvalue of $\texttt{diag}(Q, R)$, which is simple to compute.
\end{remark}

\begin{remark} \label{rem:computation:q} 
The computation of $q_1$, $q_2$, $q_3$, $z^*_3$ \eqref{eq:solve:QP3:z} and the right-hand-side of \eqref{eq:solve:QP3:mu:step:1} are not performed using the matrix multiplications shown in their expressions.
Instead, the particular structure of the matrices allows for a matrix-free computation of the vectors.
This is also true for the forward and backward substitutions performed to solve \eqref{eq:solve:QP3:mu}.
That is, we do not store all the non-zero elements of the matrices along with arrays indicating their position.
Only $\rho$ and the matrices $\alpha^i$, $\beta^i$, $(\rho Q)^{-1}$, $(\rho R)^{-1}$, $T$, $S$, $A$ and $B$ are stored.
\end{remark}

\begin{remark} \label{rem:warmstart} 
    We note that the number of iterations of Algorithm \ref{alg:EADMM:MPCT} can be reduced by the application of a warmstart procedure. This paper omits discussion on this topic because it focuses on the algorithm itself. See \cite[\S 7]{Krupa_arXiv_MPCT_20_v2} for a warmstart procedure that follows the prediction-correction warmstart from \cite{Paternain_CDC_2019}.
\end{remark}

\section{Numerical results} \label{sec:numerical:results}

The approach we present in this article for solving the MPCT formulation \eqref{eq:MPCT} decouples the decision variables into three separate optimization problems in order to attain a simple structure that can be efficiently exploited using an approach similar to the one presented in \cite{Krupa_TCST_20}.
However, there are other ways to solve \eqref{eq:MPCT}.
In this section we compare the performance of Algorithm \ref{alg:EADMM:MPCT} with some direct alternatives by applying them to the ball and plate system described in \cite[\S V.A]{Krupa_TAC_2020}.
This system consists of a plate that pivots around its center point and whose inclination can be manipulated by changing the angle of its two perpendicular axes.
The system has $8$ states and $2$ control inputs, and its control objective is to steer the ball to a given position on the plate.
The control input $u = (u_{(1)}, u_{(2)}) \in \R^2$ is the angle acceleration of each of the axes of the plate, which we constrain to $|u_{(i)}|\leq 0.2$, $i \in Z_1^2$.
We start the system with the ball placed in the central point of the plate and set the reference to the point $(1.5, 1.4)$.
For a more in-depth explanation of the system we refer the reader to \cite{Krupa_TAC_2020} and \cite{Wang_ISA_2014}.

We perform closed-loop tests of the above setup using:
\begin{itemize}
    \item Algorithm \ref{alg:EADMM:MPCT} taking $z_2^0 = 0$, $z_3^0 = 0$, $\lambda^0 = 0$, and $\rho = 40$ for the constraints listed in Remark \ref{rem:rho:matrix} and $\rho = 2$ for the rest.
        We set the exit tolerance to $\epsilon = 10^{-4}$.
    \item The OSQP solver (version \texttt{0.6.0}) \cite{Stellato_OSQP}.
        Problem \eqref{eq:MPCT} can be directly cast as a QP problem by taking the decision variables as 
        \begin{equation*}
            z = (x_0, u_0, x_1, u_1, \dots, x_{N-1}, u_{N-1}, x_s, u_s).
        \end{equation*}
        In this case, the Hessian is no longer block diagonal, which means we cannot use the approach from \cite{Krupa_TCST_20}.
        However, the QP problem is still sparse (both the Hessian and the matrix that defines the equality constraints are sparse).
        Therefore, a sparse solver for general QP problems should solve it efficiently.
        A good choice in our case is OSQP, since it is a well known QP solver based on ADMM.
        We use its default options with a few exceptions: the exit tolerances are all set to $10^{-4}$ and the warmstart procedure is disabled to better compare its ADMM algorithm with the other alternatives we list. 
    \item Extended state space: we recover a block diagonal Hessian by extending the state space with the artificial reference.
        That is, by defining $\hat x_i = (x_i, x_s)$ and $\hat u_i = (u_i, u_s)$, $i \in \Z_0^{N-1}$, problem \eqref{eq:MPCT} can be rewritten so that the resulting QP has a block diagonal Hessian at the expense of increasing the number of decision variables and constraints, including additional constraints that ensure that $x_s$ and $u_s$ are the same in all the prediction steps.
        We solve the resulting QP problem using the sparse ADMM-based solver from \cite{Krupa_TCST_20}, where the matrices are stored using standard sparse-matrix representations.
        We take $\rho = 1$ and the exit tolerance as $\epsilon = 10^{-4}$.
    \item Standard MPC: the ADMM-based solver presented in \cite{Krupa_TCST_20} for the standard MPC formulation shown in equation (9) of \cite{Krupa_TCST_20}.
        This comparison is of interest because, in terms of iteration complexity, the proposed solver is very similar to the ADMM-based solver from \cite{Krupa_TCST_20}, in that each iteration of the algorithm requires \textit{(i)} solving a system of equations \eqref{eq:solve:QP3:mu} and \textit{(ii)} computing \eqref{eq:solve:QP1}. The main difference between the two is that Algorithm \ref{alg:EADMM:MPCT} also requires the evaluation of \eqref{eq:solve:QP2} at each iteration, which is computationally cheap in comparison to  the other steps.
        Therefore, the comparison with \cite{Krupa_TCST_20} serves as an indication of the ``price-to-pay" for the inclusion of the artificial reference.
        We take $\rho = 1$ and the exit tolerance as $\epsilon = 10^{-4}$.
\end{itemize}

\begin{remark}
There are many possible efficient alternatives for solving \eqref{eq:MPCT} other than the ADMM-based approaches discussed here, including interior point methods \cite{Rao_JOTA_1998}, active set methods \cite{Saraf_TAC_2019}, and even other non-ADMM Lagrangian methods \cite{Chiang_TAC_2017}.
However, we restrict our comparison here to the use of ADMM-based approaches since our objective is to evaluate if the use of the EADMM algorithm is worthwhile when compared to similar alternatives, and not to determine if it outclasses other optimization methods.
\end{remark}

For the MPCT formulation we take $N = 30$,
\begin{align*}
    Q &= \texttt{diag}(10, 0.05, 0.05, 0.05, 10, 0.05, 0.05, 0.05),\\
    R &= \texttt{diag}(0.5, 0.5), \quad S = \texttt{diag}(0.3, 0.3), \\
    T &= \texttt{diag}(600, 50, 50, 50, 600, 50, 50, 50).
\end{align*}
For the MPC formulation from \cite{Krupa_TCST_20} we take the same $N$, $Q$, $R$ and $T$ shown above.
The tests are run on a Linux laptop (Intel i5 processor) using Matlab MEX files compiled with the \textit{gcc} compiler using the \texttt{O3} flag.
With the obvious exception of OSQP, all solvers where generated using version \texttt{v0.1.5} of the SPCIES toolbox \cite{Spcies}.
The penalty parameters were hand tuned to achieve a good performance of the solvers.

\begin{figure*}[ht]
    \centering
    \begin{subfigure}[ht]{0.47\textwidth}
        \includegraphics[width=\linewidth]{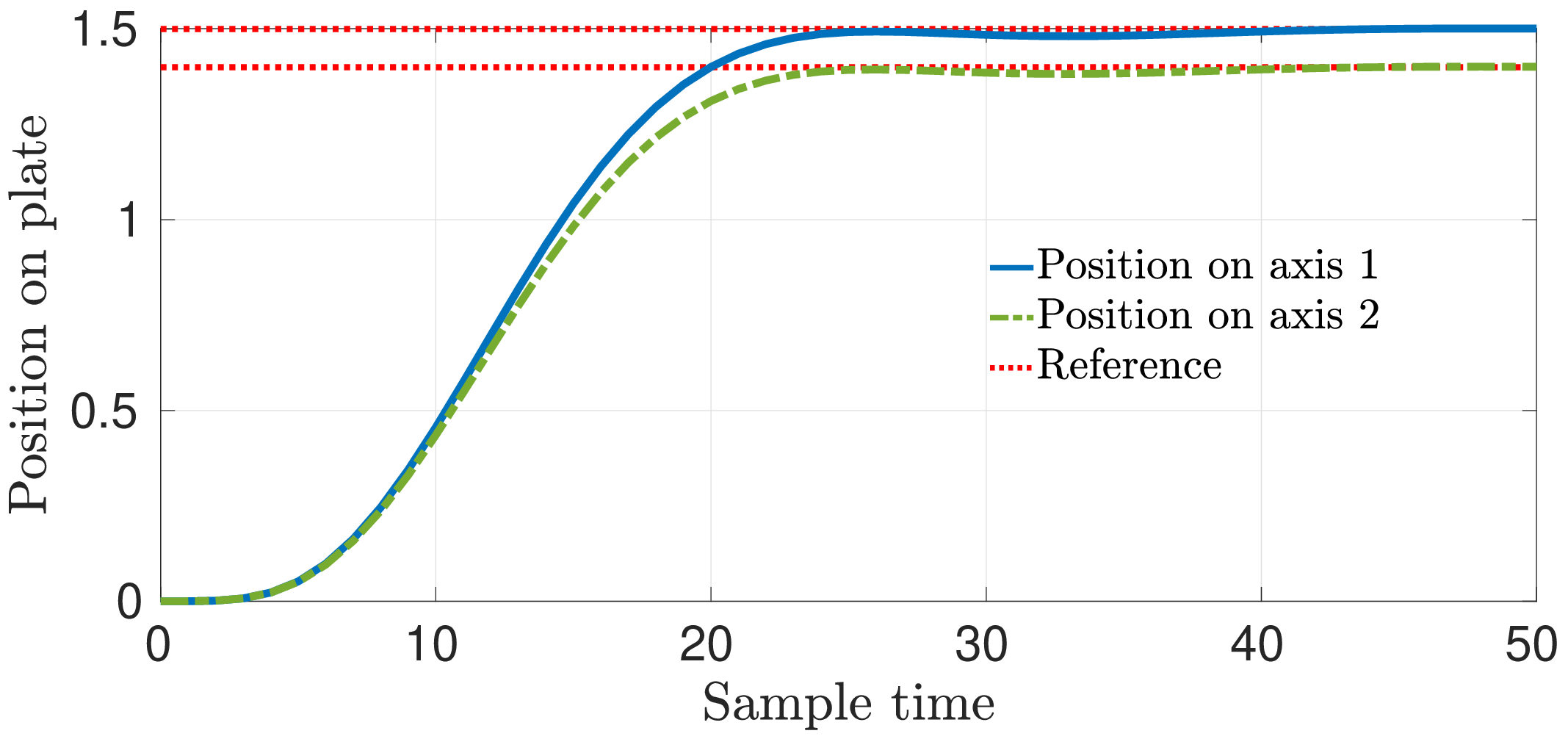}
        \caption{Position of ball for the MPCT controller.}
        \label{fig:BaP:state}
    \end{subfigure}%
    \hfill
    \begin{subfigure}[ht]{0.47\textwidth}
        \includegraphics[width=\linewidth]{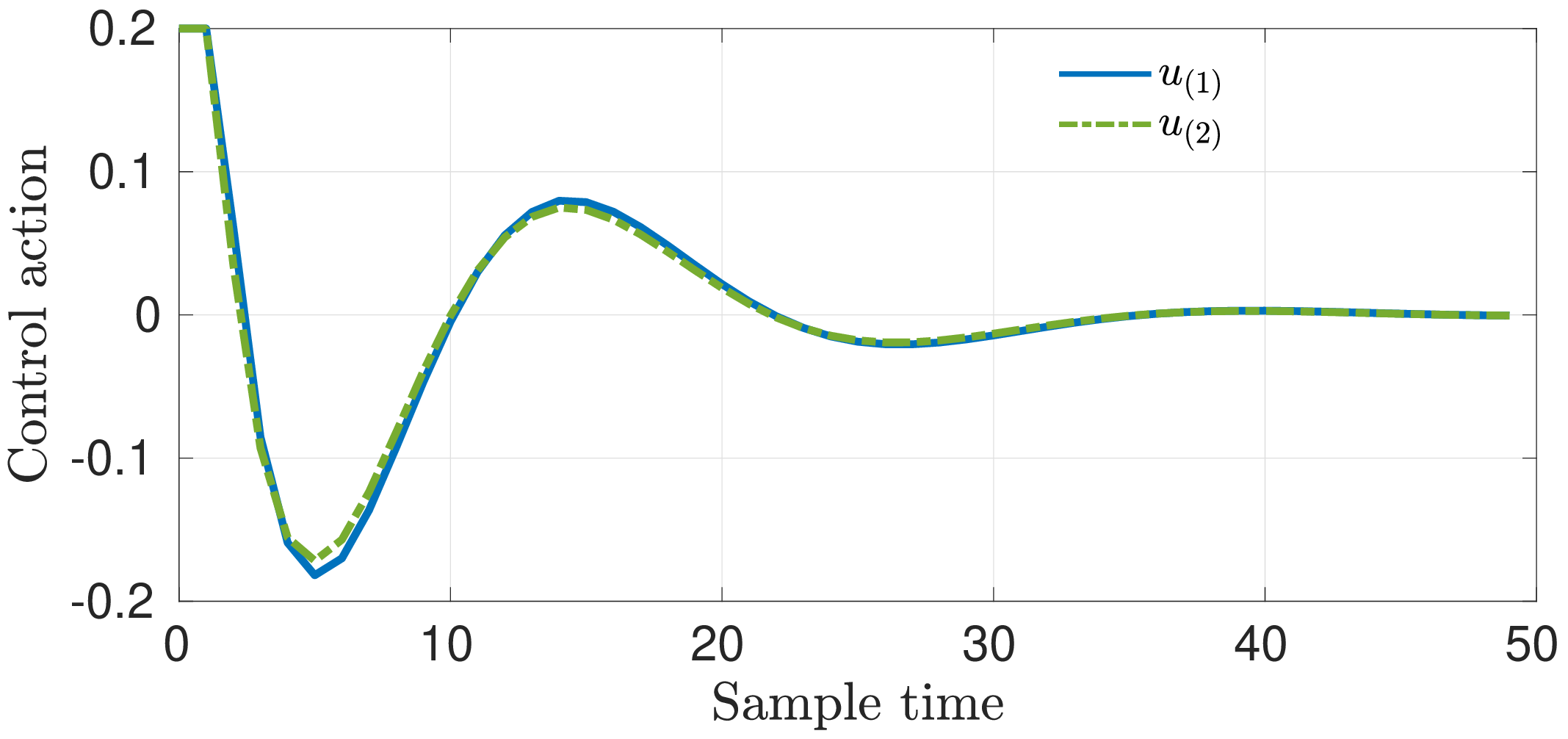}
        \caption{Control input for the MPCT controller.}
        \label{fig:BaP:input}
    \end{subfigure}%
    \hfill

    \begin{subfigure}[ht]{0.47\textwidth}
        \includegraphics[width=\linewidth]{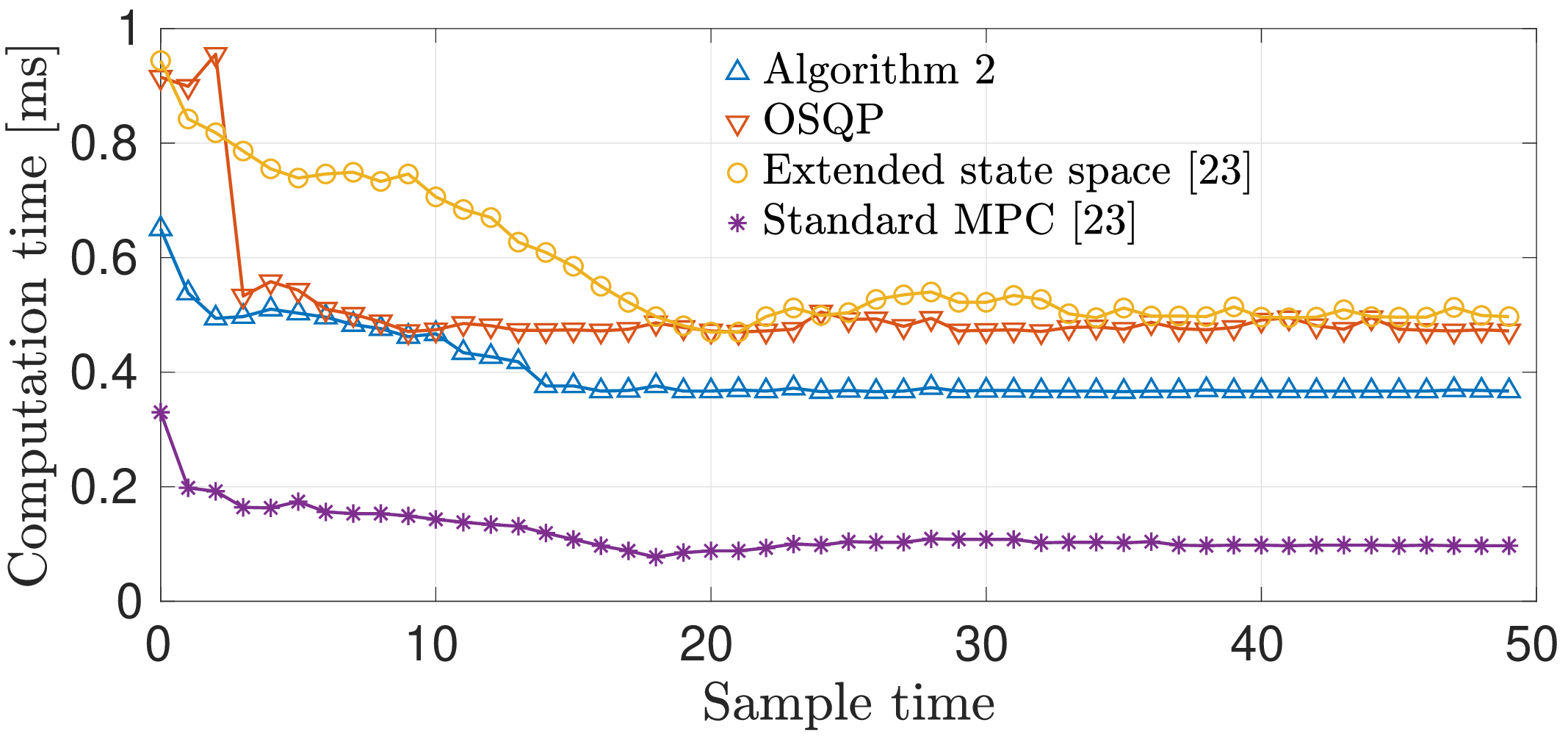}
        \caption{Computation times.}
        \label{fig:BaP:computation}
    \end{subfigure}%
    \hfill
    \begin{subfigure}[ht]{0.47\textwidth}
        \includegraphics[width=\linewidth]{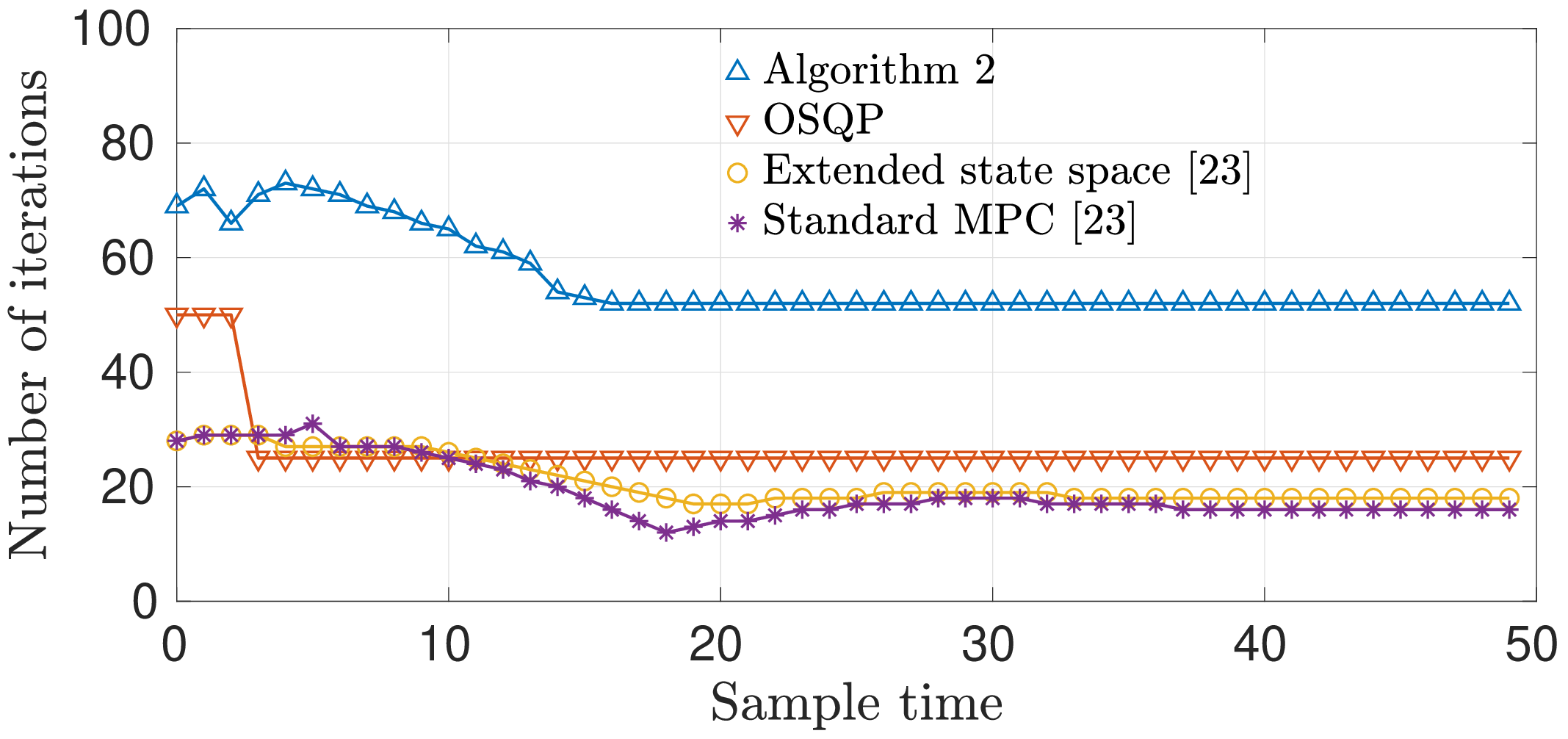}
        \caption{Number of iterations.}
        \label{fig:BaP:iter}
    \end{subfigure}%
    \caption{Ball and plate system closed-loop tests.}
    \label{fig:BaP}
\end{figure*}

Figure \ref{fig:BaP} shows the results of the tests.
Figures \ref{fig:BaP:state} and \ref{fig:BaP:input} show the trajectory of the position of the ball in each axis and the control inputs, respectively, using the MPCT controller.
The trajectories are only shown once because the results using each of the MPCT solvers are indistinguishable to the naked eye.
The trajectories for the standard MPC formulation are not shown because we are only interested in its computational results. They are very similar to the ones obtained with MPCT.
Figures \ref{fig:BaP:computation} and \ref{fig:BaP:iter} show the computation time and the number of iterations of each solver, respectively. 

The advantage of the proposed approach can be seen in its iteration complexity.
The average computation times (in milliseconds) per iteration of the results shown in Figure \ref{fig:BaP} are $0.0071$ for the proposed solver, $0.0062$ for the solver from \cite{Krupa_TCST_20} applied to the standard MPC formulation, $0.0193$ for OSQP and $0.0279$ for the extended state space approach.
That is, compared to the solver from \cite{Krupa_TCST_20}, the addition of the artificial reference only increases the computation time per iteration by $~15.1$\% (in this particular example).

However, we find that, in general, the EADMM algorithm takes more iterations to converge than the other solvers.
Even so, it can still outperform other sparse solvers due to its low iteration complexity, especially if applied to larger systems and/or with large prediction horizons, where the advantage of not storing all the non-zero elements of the matrices and of performing the matrix-vector operations by direct identification of the structures becomes more prevalent.
For smaller systems and short prediction horizons, the advantages of the small memory footprint and the iteration complexity may not outweigh the higher typical number of iterations.
In any case, the use of one approach or another would have to be determined in a case by case basis.
For instance, in the results shown in Figure \ref{fig:BaP} the proposed solver outperforms OSQP and the sparse solver based on extending the state space, in spite of it requiring more iterations.

Typically, the number of iterations of first order methods can become large if there are active constraints in the optimal solution.
As shown in Figure \ref{fig:BaP:input}, this was not the case during the first few sample times, in spite of the control actions reaching their upper bound.
However, this phenomenon may occur if there are additional active constraints along the prediction horizon (particularly so if there are active state constraints).
This is true for the proposed solver as well as for any other ADMM-based approach.

\begin{figure}[t]
    \includegraphics[width=0.98\linewidth]{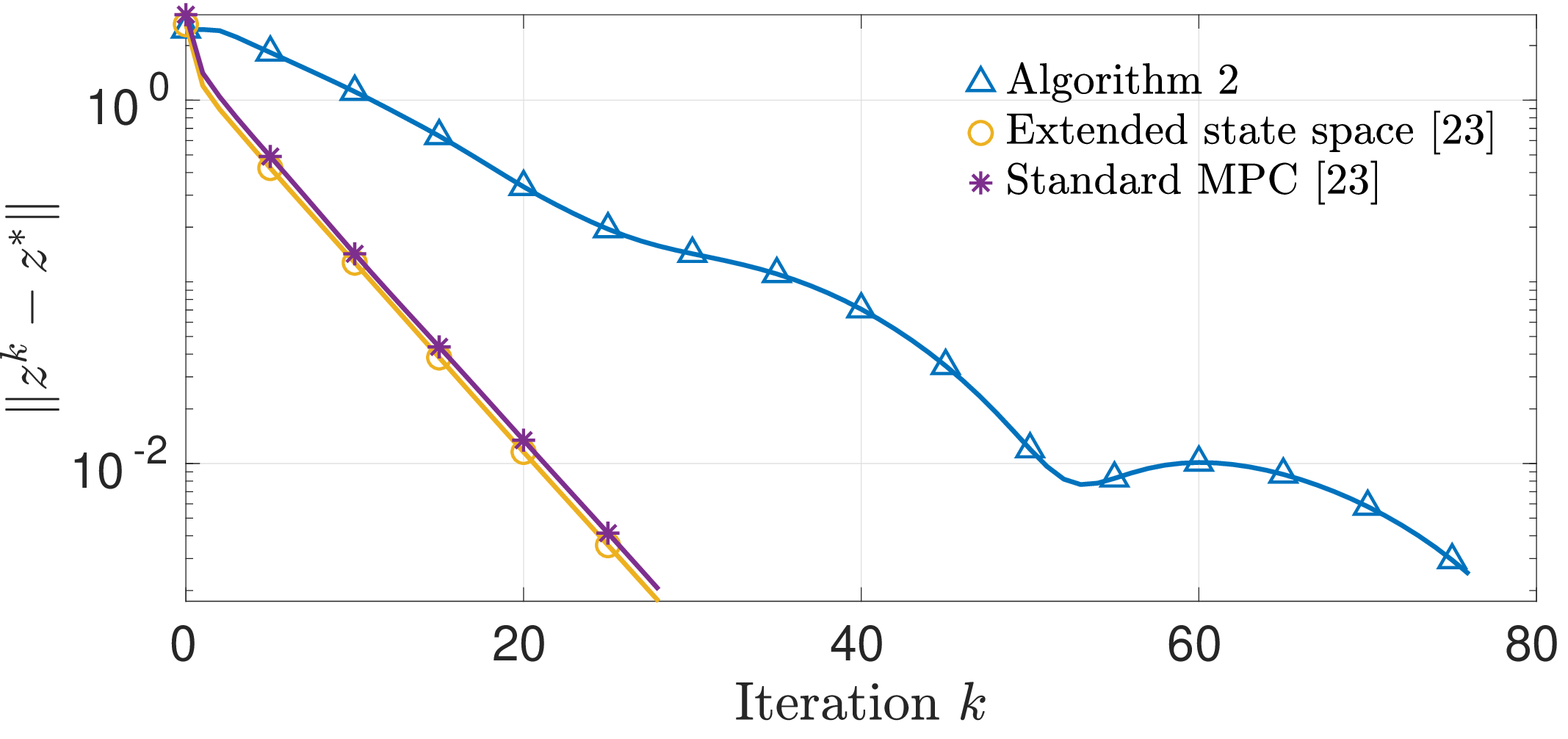}
    \caption{Convergence of the SPCIES solvers.}
    \label{fig:convergence}
\end{figure}

Figure \ref{fig:convergence} shows the convergence of the decision variables of the solvers obtained from the SPCIES toolbox to the optimal solution of their QP problems during the first sample time of the results shown in Figure \ref{fig:BaP}.
For Algorithm \ref{alg:EADMM:MPCT}, decision variable $z_1^k$ is used to compute the distance.
The optimal solutions of the QP problems are obtained using the \texttt{quadprog} solver from Matlab.
The results show that the convergence of the EADMM algorithm is non-monotone and slower than the ones obtained using ADMM.
Non-monotonicity is a phenomenon found in some classes of first order methods \cite{Alamo_Restart_CDC_19}.

Further evidence of the good performance of the proposed solver can be found in \cite[\S 5.8]{Krupa_Thesis_21} and in \cite{Krupa_ECC_21_arXiv}, which shows its implementation in a Raspberry Pi to control a real two-wheeled inverted pendulum robot in real-time.

\section{Conclusions} \label{sec:conclusions}

This article presents a sparse solver for the linear MPC for tracking formulation based on the extended ADMM algorithm.
We show how the use of this method, along with an appropriate  selection of decision variables, leads to an efficient solver with simple matrix structures that can be exploited to attain a small iteration complexity and memory footprint.
This statement is supported by numerical results comparing the proposed solver to other ADMM-based approaches, which suggest that, from a computational point of view, the proposed approach can provide better results.
The solver is available in the SPCIES toolbox for Matlab \cite{Spcies} at \url{https://github.com/GepocUS/Spcies}.


\bibliographystyle{IEEEtran}
\bibliography{IEEEabrv,BibKrupa}

\end{document}